\documentclass[12pt,a4paper]{article}

\newcommand{\be}{\begin{equation}}
\newcommand{\ee}{\end{equation}}
\newcommand{\bea}{\begin{eqnarray}}
\newcommand{\eea}{\end{eqnarray}}

\begin{document}

\title{Path Integral Formulation of Noncommutative Quantum Mechanics}

\author{Ciprian Acatrinei\thanks{ e-mail: acatrine@physics.uoc.gr.} \\
             \\
        {\it Department of Physics, University of Crete,}\\
        {\it P.O. Box 2208, Heraklion, Greece} \\
        and \\
        {\it National Institute of Nuclear Physics and Engineering}\\
        {\it P.O. Box MG-6, Bucharest, Romania}   }

\date{10 July, 2001}

\maketitle

\begin{abstract}
We propose a phase-space path integral formulation of noncommutative
quantum mechanics, and prove its equivalence to the operatorial formulation.
As an illustration, the partition function of a noncommutative two-dimensional
harmonic oscillator is calculated.
\end{abstract}

Noncommutative quantum mechanics represents a natural extension
of usual quantum mechanics, in which one allows nonvanishing
commutators  also between the coordinates, and between the momenta.
Denoting the coordinates and the momenta collectively  by $\{x_i\}$,
in (2+1)-dimensions one has the commutation relations
\be
\label{cr}
[x_i,x_j]=i \Theta_{ij}, \quad x_{1,2,3,4}=q_1,q_2,p_1,p_2 ,
\ee
where the antisymmetric matrix $\Theta_{ij}=(\omega^{-1})_{ij}$ is given by
\be
\label{setup}
\Theta=
\left (
\begin{array}{rrrr}
0 & \theta & 1 & 0  \\
-\theta & 0 & 0 &1 \\
-1 &  0 & 0 & \sigma \\
0 & -1  & -\sigma &  0
\end{array}
\right )
\quad \mbox{i.e.} \quad
\omega=\frac{1}{1-\theta\sigma}
\left (
\begin{array}{rrrr}
0 & \sigma &-1 & 0  \\
-\sigma & 0 & 0 &-1\\
1 &  0 & 0 &  \theta\\
0 & 1  &  -\theta & 0
\end{array}
\right ).
\ee


The purpose of this note is to provide
a path integral formulation of quantum dynamics,
which is consistent with
(\ref{cr}) and (\ref{setup}).
We will propose a phase-space path integral, and derive from it the
commutation relations (\ref{cr}) and the extended Heisenberg equations
of motion. Conversely, we will start from the operatorial formalism
and derive the above mentioned path-integral.
As an illustration, the partition function of a two-dimensional
harmonic oscillator will be calculated, and compared to the commutative
case.
Although we work in a (2+1)-dimensional space-time,
our considerations can be easily extended to higher dimensional spaces.

An interesting path integral approach to noncommutative
(\`{a} la Connes) spaces was given in \cite{gianpiero}.
Some classical aspects of noncommutative mechanics were described
in \cite{eu}. Noncommutative quantum mechanics generated much
interest recently: a partial list of references can be found in \cite{ncqm};
attempts to introduce path integrals appeared in \cite{attempts}.
For some earlier work relevant to our topic, see \cite{peierls_1}
and references therein.

Let us start from the classical action
\be
S=\int_0^T d t
\left ( \frac{1}{2}\omega_{ij}x_i \dot{x}_j - H(x) \right ),
\quad x_{1,2,3,4}=q_1,q_2,p_1,p_2,  \label{action}
\ee
where
$\omega=(\Theta)^{-1}$, and $H$ is the Hamiltonian of the system.
The Hamilton equations of motion
(which are derived through variation of (\ref{action}),
under the requirement $\delta x_i\left|_{0,T}=0\right.,i=1,2,3,4$)
and the basic Poisson brackets are
\be
\dot{x}_i=\{x_i,H\}=\Theta_{ij}\frac{\partial H}{\partial x_j},
\qquad
\{x_i,x_j\}=\Theta_{ij}.   \label{em0}
\ee
Above,
the extended Poisson bracket is
$\{A,B\}\equiv
\frac{\partial A}{\partial x_i}\Theta_{ij}\frac{\partial B}{\partial x_j}$.

For
nonlinear systems, i.e.
higher than quadratic actions,
and if $\theta \neq 0$, equations (\ref{em0})
do not admit a Lagrangian formulation \cite{eu}.
Thus, in general, one can at best hope for a phase-space
path integral formulation of the quantum theory corresponding to the action
(\ref{action}).

This is provided
by the path integral
\be
Z=\int\prod_{k=1}^{4} D x_k e^{i S}=\int\prod_{k=1}^{4} D x_k e^{i
\int dt \left ( \frac{1}{2}\omega_{ij}x_i \dot{x}_j - H(x) \right )}.
\label{pi}
\ee
The prescription (\ref{pi}) is simple:
if $[\hat{x}_i,\hat{x}_j]=i\Theta_{ij}$
then
$Z=\int Dx e^{i\int dt (\Theta^{-1}_{ij}\frac{x_i\dot{x}_j}{2}-H)}$,
and general: it applies to any Hamiltonian $H$.
The physical meaning of the above path integral
will become clear when we will derive it from the canonical formalism.
To find the commutation relations and the operatorial equations of motion
enforced by (\ref{pi}) - our first task,
all we need to know is that
$Z$ represents a transition amplitude between two states of a given Hilbert space,
and that
time-ordering of operators is enforced, as usual, by the path integral,
$\int Dx O_1 O_2 e^{iS}=\langle T\{\hat{O}_1 \hat{O}_2\} \rangle$.

To derive in an elementary way the commutation relations enforced by (\ref{pi}),
we discretize the path integral:
\be
Z\simeq \int\prod_{(n)=1}^{N}\prod_{k=1}^{4} D x_k^{(n)}e^{i
\sum_{(n)} \left ( \frac{1}{2}\omega_{ij}x_i^{(n)} (x_j^{(n+1)}-x_j^{(n-1)})
- \epsilon H(x_i^{(n)}) \right )};
\label{discrete_pi}
\ee
$\epsilon$ is the time increment, $\epsilon=T/N$, and $x_i^{(n)}$ is the value
of the phase space variable $x_i$ at time $t_0+n\epsilon$, $n=0,1,2,\dots,N$.
Let us consider the expectation value of $\frac{\partial \hat{O}}{\partial \hat{x}_k^{(n)}}$,
where $\hat{O}(\hat{x}_i)$ is an operator depending on the $\hat{x}_i$'s.
Integrating by parts under the path integral (\ref{discrete_pi}), one gets
\be  \label{master}
\left < \frac{\partial \hat{O}}{\partial \hat{x}_k^{(n)}}\right > =
-i\left < T \{ \hat{O}(\hat{x}_i),
\frac{\partial \tilde{S}}{\partial \hat{x}_k^{(n)}}\right  > .
\ee
$T\{,\}$ represents the time-ordering
of operators, which means $(n)$-ordering in the discrete case.
$\tilde{S}$  is the discretized form of the action, and
$
\frac
{  \partial \tilde{S}  }
{  \partial \hat{x}_k^{(n)}  }
=
\omega_{kj}  (   \hat{x}_j^{(n+1)}  -  \hat{x}_j^{(n-1)}   )
-\epsilon
\frac
{  \partial \hat{H}   }
{  \partial \hat{x}_k^{(n)}   }
$.
Choosing $\hat{O}=\hat{x}_i$,
converting the  $(n)$-ordering into a commutator,
and taking then the continuum limit
$\epsilon \rightarrow 0$,
(\ref{master}) becomes
\be
\sum_j\omega_{ij}[\hat{x}_j,\hat{x}_k]=i\delta_{ik},
\ee
which implies
$[\hat{x}_i,\hat{x}_j]=i\Theta_{ij}=i(\omega^{-1})_{ij}$,
i.e.
the commutation relations (\ref{cr}).

To derive the Heisenberg equations of motion as well, we choose
$\hat{O}$ proportional to the identity operator. Then, one gets
$\frac{\partial \tilde{S}}{\partial \hat{x}_k^{(n)}}=0$,
leading to  $\omega_{ij}\frac{d}{dt}\hat{x}_j=\frac{\partial \hat{H}}{\partial \hat{x}_j}$.
Solving for $\frac{d}{dt}\hat{x}_i$, one obtains
\be   \label{em}
\frac{d}{dt}\hat{x}_i= \Theta_{ij}\frac{\partial \hat{H}}{\partial \hat{x}_j}
=-i[\hat{x}_i,\hat{H}],
\ee
which are the extended Heisenberg equations of motion
(the quantum form of (\ref{em0}), with $\{,\}_{PB}\rightarrow -i[,]$).

Having derived the commutation relations and the operator equations of
motion from the path integral,
we  now proceed to do the opposite,
in order to show the equivalence between the path integral
and the canonical formulations.

To work within the canonical formalism, we need
a complete basis in the Hilbert space of the theory.
Since $[\hat{p}_i,\epsilon_{ij}\hat{q}_j]=0$,
such a basis is provided, for instance, by
the set of eigenvectors of $\hat{q}_1$ and $\hat{p}_2$,
$\{|q_1,p_2>\}$, or alternatively by $\{|q_2,p_1>\}$.

We  want to calculate the transition amplitude
\be
{\cal A}=<q_1,p_2|e^{-i \hat{H} T}|q_1^{(0)},p_2^{(0)}>, \label{ta}
\ee
between two states with prescribed position along the first axis of coordinates,
and well defined  momentum along the second axis.
In what follows, we will  need $ <q_1,p_2|q_2,p_1> $.
To evaluate it, let us first notice that (\ref{cr}) implies
\bea
&e^{i\beta\hat{p}_1} |q_1,p_2>&=|q_1-\beta,p_2+\beta \sigma>  \\
&e^{i\gamma\hat{q}_2}|q_1,p_2>&=|q_1-\gamma\theta,p_2+\gamma>.
\label{global}
\eea
On a wave function $\psi(q_1,p_2)=<q_1,p_2|\psi>$,
the operators  $\hat{p}_1$ and $  \hat{q}_2$ will consequently act as follows:
\be
\hat{p}_1=\frac{1}{i}(\partial_{q_1}-\sigma \partial_{p_2})
\qquad
\hat{q}_2=\frac{1}{i}(\theta \partial_{q_1}-\partial_{p_2}).
\label{local}
\ee
This implies
\bea
q_2<q_1,p_2|q_2,p_1>=&i(\partial_{p_2}-\theta \partial_{q_1})&<q_1,p_2|q_2,p_1>  \\
p_1<q_1,p_2|q_2,p_1>=&i(-\partial_{q_1}+\sigma \partial_{p_2})&<q_1,p_2|q_2,p_1>,
\eea
and we end up with
\be
<q_1,p_2|q_2,p_1>=\frac{1}{2}
\exp\left(\frac{i}{1-\theta\sigma}(q_1 p_1-q_2 p_2+\theta p_1 p_2-\sigma q_1 q_2)\right).
      \label{bracket}
\ee
We can now proceed to derive a path integral expression for (\ref{ta}).
For simplicity, we consider Hamiltonians of the form
$H=\frac{1}{2m}(p_1^2+p_2^2)+V(q_1,q_2)$.
(A magnetic field $\sigma$ is already present due to
$[p_1,p_2]=i\sigma\neq 0$).
Since what follows is a variation of the standard procedure,
only the main steps needed to calculate (\ref{ta}) will be mentioned.
First, we divide the time interval $T$ into $N+1$ short intervals, such that terms of the order
$(T/N)^2$ are negligible. Then, we insert $N$ unity operators of the form
\be
\int dq_1^{(n)} \int dp_2^{(n)}
|q_1^{(n)},p_2^{(n)}><q_1^{(n)},p_2^{(n)}|, \quad (n)=1,2\dots ,N,
\ee
followed by other $N+1$ unity operators
\be
\int dq_2^k \int dp_1^k
|q_2^{(n)},p_1^{(n)}><q_2^{(n)},p_1^{(n)}|,
\quad (n)=1,2\dots ,N+1.
\ee
At this stage, there are no more operators in (\ref{ta}), and one uses (\ref{bracket}).
Taking the continuum limit results in
\be
<q_1,p_2|e^{-i \hat{H} t}|q_1^{(0)},p_2^{(0)}> =\int D q_1Dq_2Dp_1Dp_2
e^{i\int dt(\frac{p_1\dot{q}_1-q_2\dot{p}_2+\theta p_1\dot{p}_2-\sigma q_2\dot{q}_1}{1-\theta\sigma}
-H)}
\ee
which, after integration by parts in the action term, is precisely (\ref{pi}).
We have thus found a natural path integral formulation of quantum mechanics on noncommutative phase-space.

Several remarks are in order:
First,
one should notice the easiness with which noncommutativity is introduced at the level of the path integral.
The additional quadratic couplings among phase space variables would not complicate substantially
the evaluation of a noncommutative partition function, once the commutative case is under control.
In contrast, nonzero $\theta$ and $\sigma$ enter in a somehow more complicated way at the level of the
equations of motion (\ref{em}).

Second,
we assumed that $[\hat{p}_i,\epsilon_{ij}\hat{q}_j]=0$. If one allows
nonvanishing commutators $[\hat{q}_1,\hat{p}_2]$ and  $[\hat{q}_2,\hat{p}_1]$ too,
one can easily show that
all the four operators $\hat{x}_i$,
which now act on a reduced Hilbert space spanned by the eigenvectors of only {\it one} of them,
are proportional to each other. This contradicts the commutation relations.
Consequently, we did not consider
such nonvanishing commutators in the previous analysis.

Finally, an important particular case appears when the matrix $\Theta$ is singular,
namely when $\theta\sigma=1$.
In this case,
\be  \label{reduction}
\hat{q}_1=-\theta \hat{p}_2   \qquad    \hat{q}_2=\theta \hat{p}_1
\ee
and the initial two-dimensional (2D) problem reduces to a 1D one.
The price to pay is, in general, that the 1D Hamiltonian
will be nonlinear in both the coordinates and the momenta.
The dimensionally reduced path integral is,
in the notation $q=q_1=-\theta p_2$, $p=p_1=(1/\theta) q_2$,
\be
\int Dq Dp e^{ i\int dt   ( p \dot{q}-\frac{p^2}{2m}-\frac{q^2}{2m\theta^2}-V(q,\theta p) )  }.
\ee
$V$ has the same functional form as the initial 2D potential $V(q_1,q_2)$.
A (singular) noncommutative theory living in two space dimensions
is equivalent to a particular commutative one living in
one spatial dimension.

Let us
illustrate our considerations with an example, namely the partition function
of a 2D noncommutative harmonic oscillator of mass $m$ and frequency $\omega$.
The Hamiltonian is
$H_{h.o.}=\frac{p_1^2+p_2^2}{2m}+\frac{m\omega^2}{2}(q_1^2+q_2^2)$,
and we want to evaluate, for nondegenerate $\omega$, the partition function
\be
{\bf z}=\int_{x_k(0)=x_k(T)}
\prod_k D x_k e^{i\int_0^T dt (\omega_{ij}x_i\dot{x}_j/2-H_{h.o.})}.
\ee
We consider periodic boundary conditions, $x_i(0)=x_i(T)$, for simplicity.
Developing in Fourier modes,
\be
x_i(t)=\sum_n x_i^{(n)} e^{2\pi n t/T},
\ee
one can reexpress the action as a function of the $x_i^{(n)}$'s, then integrate on the $x_i^{(n)}$'s,
to obtain ${\bf z}\sim \prod_n \Delta_n^{-1}$, where
\be
\Delta_n=\det
\left (
\begin{array}{rrrr}
-m\omega^2 & i\alpha n\theta &  i\alpha n & 0 \\
-i\alpha n\theta &   -m\omega^2 &    0   &  i\alpha n  \\
- i\alpha n &  0 & -1/m &  i\alpha n\sigma \\
0 & - i\alpha n & - i\alpha n\sigma  &  -1/m
\end{array}
\right ),
\qquad
\alpha=\frac{2\pi}{(1-\theta\sigma)T}.
\ee
Obviously, when $\theta=0$ and $\sigma=0$, ${\bf z}$ becomes the usual, commutative, partition function,
which we denote by ${\bf z_0}$.
The easiest way to proceed is to compare ${\bf z}$ and ${\bf z_0}$.
Due to the Gaussian integrations,
their ratio is given by
\be
\frac{{\bf z}}{{\bf z_0}}=\prod_{n=1}^{\infty}\frac{\Delta_n^0}{\Delta_n},
\qquad
\Delta_n^0\equiv\Delta_n(\theta=0,\sigma=0).
\label{ratio}
\ee
While evaluating (\ref{ratio}), one encounters a divergent infinite product,
which we zeta-function regularize, via
$\sum_{k\geq 1}1=\lim_{s\rightarrow 0}\zeta (s)=-\frac{1}{2}$:
$\prod_{n=1}^{\infty}(1-\theta\sigma)^2\rightarrow (1-\theta\sigma)^{-1}$.
Finally,
using $\prod_{n=1}^{\infty}(1-\frac{a^2}{n^2})=\frac{\sin(\pi a)}{\pi a}$,
one obtains
\be
 {\bf z}=\frac{{\bf z_0}}{(1-\theta\sigma)} \times
  \frac{\sin^2(\omega T/2)}{\sin(\omega_1 T/2 )\sin(\omega_2 T/2)}.
\ee
Apart from the phase-space volume factor $\frac{1}{(1-\theta\sigma)}$,
noncommutativity just trades the isotropic oscillator for an anisotropic one, with frequencies
$\omega_1=\omega A_1$ and $\omega_2=\omega A_2$.
The correction factors $A_{1,2}$ are
\be
A_{1,2}=1+\frac{1}{2}((\theta m \omega)^2+(\frac{\sigma}{m\omega})^2)\pm\frac{1}{2}
(\theta m \omega+\frac{\sigma}{m\omega})^2\sqrt{4+(\theta m \omega-\frac{\sigma}{m\omega})^2}.
\ee

If $\omega =0$ (free noncommutative particle) then
$\theta$ plays no role, and
\be
{\bf z}(\omega=0)={\bf z_0}(\omega=0) \times
\frac{(\frac{T\sigma}{2m})}{\sin(\frac{T\sigma}{2m})},
\ee
which is the partition function of a 2D particle in a magnetic field $\sigma$.

If $\theta\sigma=1$,
it is easily seen that
an isotropic 2D noncommutative harmonic oscillator  reduces to a
1D commutative one, with mass $M=\frac{m}{1+\theta^2\omega^2m^2}$ and frequency
$\Omega=\frac{1}{\theta M}$ .

\subsection*{Acknowledgements}
I am grateful to Gianpiero Mangano for fruitful discussions.
I thank the HEP theory group of the University of Crete for kind hospitality,
and MURST (Italy) for financial support.

\end{document}